\documentclass[final,5p,times,twocolumn,authoryear]{elsarticle}

\usepackage{amssymb}
\usepackage{amsmath, nccmath}

\usepackage{url}
\usepackage[dvipsnames]{xcolor}
\usepackage{verbatim}
\usepackage{hyperref}

\journal{Astronomy and Computing}

\begin{document}
\newcommand{\bmat}[1]{\mathrm{\textbf{#1}}}
\newcommand{\bvec}[1]{\textit{\textbf{#1}}}
\newcommand{\de}{\mathrm{d}}
\newcommand{\lapack}{\texttt{LAPACK}}
\newcommand{\magma}{\texttt{MAGMA}}
\newcommand{\mpi}{\texttt{MPI}}
\newcommand{\nptm}{\texttt{NP\_TMcode}}
\newcommand{\omp}{\texttt{OpenMP}}
\newcommand{\glm}[1]{\textcolor{ForestGreen}{\emph{(GLM:)} #1}}
\newcommand{\gmu}[1]{\textcolor{red}{\emph{(GMu:)} #1}}
\newcommand{\gba}[1]{\textcolor{blue}{\emph{(gba:)} #1}}
\definecolor{amethyst}{rgb}{0.6, 0.4, 0.8}
\newcommand{\rsj}[1]{\textcolor{amethyst}{\emph{(RSj:)} #1}}
\definecolor{amber}{rgb}{1.0, 0.75, 0.0}
\newcommand{\mai}[1]{\textcolor{amber}{\emph{(MAI:)} #1}}
\definecolor{awesome}{rgb}{1.0, 0.13, 0.32}
\newcommand{\ccp}[1]{\textcolor{awesome}{\emph{(CCP:)} #1}}
\definecolor{chocolate(traditional)}{rgb}{0.48, 0.25, 0.0}
\newcommand{\shr}[1]{\textcolor{chocolate(traditional)}{\emph{(ShR:)} #1}}


\begin{frontmatter}

\title{Nano-particle Transition Matrix code implementation} 

\author[inaf-oaca]{G. La Mura} 
\author[inaf-oaca]{G. Mulas} 
\author[inaf-oaca]{G. Aresu} 
\author[cnr-ipcf]{M. A. Iat\`{i}} 
\author[inaf-oapa]{C. Cecchi-Pestellini} 
\author[unime-fis,cnr-ipcf,uni-kur]{S. Rezaei} 
\author[unime-fis]{R. Saija} 

\affiliation[inaf-oaca]{organization={INAF - Astronomical Observatory of Cagliari},
            addressline={Via della Scienza 5}, 
            city={Selargius},
            postcode={09047}, 
            state={CA},
            country={Italy}}

\affiliation[cnr-ipcf]{organization={CNR - Institute for Chemo-Physical Processes},
                addressline={Viale F. Stagno d'Alcontres 37},
                city={Messina},
                postcode={98158},
                country={Italy}}

\affiliation[inaf-oapa]{organization={INAF - Osservatorio Astronomico di Palermo},
                addressline={Piazza del Parlamento 1},
                city={Palermo},
                postcode={90134},
                country={Italy}}

\affiliation[unime-fis]{organization={University of Messina - Dep. of Mathematical and Computer Sciences, Physical Sciences and Earth Sciences},
            addressline={Viale F. Stagno D'Alcontres 31}, 
            city={Messina},
            postcode={98166}, 
            state={ME},
            country={Italy}}

\affiliation[uni-kur]{organization={University of Kurdistan,Dept. of Physics, Faculty of Science},
                addressline={Pasdaran Street},
                city={Sanandaj},
                postcode={416/6613566176},
                country={Iran}}

\begin{abstract}
  Electromagnetic scattering and absorption by material particles is a fundamental physical problem with a broad range of applications, going from laboratory experiments, biology and material sciences, all the way up to environmental studies and astrophysical investigations. In spite of its primary importance, an exact theoretical treatment is only possible for a limited range of ideal cases, while realistic situations require the development of numerical solutions. In the course of the years, several techniques were developed to model the effects of scattering and absorption in more general cases, using approaches such as the Discrete Dipole Approximation (DDA), the Finite Difference Time Domain (FDTD) method, the Transition Matrix formalism (T-matrix) or the Mean Field Theory (MFT). Among these possibilities, the T-matrix approach grants the highest degree of flexibility in modeling aggregates of spherically symmetric particles with arbitrary  overall morphology and composition, but its application has been limited by the challenging computational requirements of the method in realistic cases. This paper describes the Nano-Particle Transition Matrix Code project (\nptm), a new implementation of the T-matrix formalism that, taking advantage of high performance parallel hardware architectures, allows the solution of increasingly complex models, while substantially reducing the computing time. The paper describes the code structure, with a particular focus on the algorithm optimization, and it presents the results of the performance analysis for a set of development applications.
\end{abstract}




\end{frontmatter}



\section{Introduction}
\label{sec-intro}

Understanding the propagation of electromagnetic radiation in non homogeneous transmissive media represents a critical question that is acquiring a growing role in several fields of scientific research and technological development. While the average properties of radiation propagation can be easily described in all scales where the traversed medium may be regarded as homogeneous, the presence of inhomogeneities, such as distributions of material particles with different shapes and compositions, requires a much higher level of detail to be properly interpreted. This problem is well known in Astrophysics, since the combination of scattering and absorption of radiation by interstellar dust grains gives raise to the process of extinction. With the advent of a new generation of spectroscopic instruments, which opened the field of exoplanetary atmosphere exploration by means of transmission spectroscopy, the issue became even more important, due to the fundamental role played by suspended particles in the form of aerosols in determining the transmittance of atmospheres at different wavelengths \citep{Gao21}. In all cases, the efficiency with which material particles can absorb, scatter and re-emit radiation is fundamental to connect the spectral signatures of various chemical species with the amount of material necessary to produce them.

In general, the interaction of radiation with material particles can be described in different regimes, depending on the ratio of the characteristic particle dimension $d$ to the radiation wavelength $\lambda$. When $\lambda \ll d$, the problem can be addressed in the framework of geometric optics. Conversely, if $\lambda \gg d$, we enter the Rayleigh approximation regime, which implies a scattering cross-section dependence on wavelength of the form $\sigma_{sca} \propto \lambda^{-4}$. Ultimately, whenever $\lambda \sim d$, a full description of the process, involving the solution of the radiation field equations, is necessary.

The theoretical treatment of interaction between radiation and matter particles having comparable size with respect to the radiation wavelength, was first addressed in the framework of the \textit{Mie theory} \citep{Mie1908}, which provides an exact description of the scattering process by solving the radiation field equations for the case of particles with rigorous spherical symmetry. Although the theory was subsequently generalized to some other specific geometries, such as, e.g., cylindrical or ellipsoidal particles, the results obtained in this way are not generally able to reproduce the properties of realistic scatterers, like dust grains, soot particles and aerosol droplets \citep{Lodge24}. For these cases, instead, numerical techniques are commonly introduced that decompose the problem down to different levels of detail, deriving partial solutions which are subsequently combined in a global description, thanks to the linearity of the radiation field equations. Such techniques can be subdivided in various broad categories, depending on the method that is used to model the involved particles. The main ones include, in particular, the Discrete Dipole Approximation \citep[DDA,][]{Purcell73,Draine94}, which represents the particles as a finite collection of contiguous electric dipoles, the Mean Field Theory (MFT), originally laid out by \citet{Berry86} and \citet{Botet97}, then further developed in the Modified Mean-field Theory (MMFT) by \citet{Tazaki18}, which can instead describe fractal aggregates of identical spheres, the Finite Difference Time Domain (FDTD) method based on the numerical integration of the Maxwell equations in the time domain using a finite-difference scheme with space and time discretizations \citep{Taflove05, Yee96}, and the Transition Matrix formalism \citep[T-matrix,][]{Waterman71,Mishchenko96,Borghese07}, which can treat arbitrary combinations of spherically symmetric particle components.

While each of the above numerical approaches has its specific combination of advantages and drawbacks, which may render one solution more convenient than others for a particular case, the T-matrix formalism is regarded as the most appropriate approach when the problem combines structural particle complexity with the necessity to take into account many different particle orientations with respect to the surrounding radiation fields. This distinguishing feature, however, is counterbalanced by the larger upfront computational cost with respect to faster partial solutions. In recent times, however, the development and diffusion of more and more efficient parallel hardware architectures allowed for the optimization of computationally intensive tasks, leading to an overall reduction of the amount of time required to solve increasingly complicated problems.

In this paper, we present a novel implementation of the T-matrix formalism, based on the framework originally developed by Borghese, Denti \& Saija, named \textit{Nano-Particle Transition Matrix code} (\nptm), which takes advantage from multi-process and multi-threaded hardware architectures to provide a scalable application suite, able to work on personal workstations as well as on computing farms, with the aim of solving increasingly complex particle models with greatly improved performance. The paper is structured as follows: in section~\ref{sec-theory}, we describe the theoretical and numerical framework used by the code; in section~\ref{sec-code}, we introduce the code application suite and discuss its optimized algorithmic implementation; in section~\ref{sec-perf}, we provide a performance analysis for a set of development cases; in section~\ref{sec-discussion} we discuss the code performance and its application range and, finally, in section~\ref{sec-conclusions} we present our concluding remarks.

\section{Theoretical framework}
\label{sec-theory}

\subsection{Spherical case: the \textit{Mie} theory}
\label{sec-mie}

The description of the interaction between electromagnetic radiation and matter is generally drawn from the combination of a set of fundamental physical processes that involve radiation absorption, emission and scattering operated by the medium that is traversed by the radiation beam. Since absorption subtracts energy from a specific frequency of the radiation beam and scattering removes it from the beam's propagation direction, their combined effect is also referred to as \textit{extinction} of radiation, a fundamental concept in astrophysics and in many fields of applied optics. A convenient way to investigate such processes is to define a set of fundamental parameters that summarize their magnitude. The best way to describe processes that intercept energy from an incident radiation field is to introduce the scattering and the absorption cross-sections $\sigma_{sca}$ and $\sigma_{abs}$, which represent the \textit{area} that a material particle presents to an incoming flux of radiation, respectively, as an absorber and as a scatterer, together with their combination in the extinction cross-section $\sigma_{ext}$ (with $\sigma_{ext}= \sigma_{sca} + \sigma_{abs}$).

Starting from the assumption of a monochromatic planar wave impinging on a spherically symmetric particle, \citet{Mie1908} derived theoretical expressions for the relevant cross-sections of the particle. Defining the particle's size parameter:
\begin{equation}
    x = \dfrac{2 \pi \sqrt{\varepsilon}}{\lambda}r_{sph}, \label{eq_size_par}
\end{equation}
where $r_{sph}$ is the radius of the particle, $\lambda$\ the wavelength of the incident radiation and $\varepsilon$ the dielectric constant of the environment medium (here assumed to be non conductive), we can introduce the \textit{Mie coefficients}:
\begin{subequations}
\begin{align}
    a_n & = \dfrac{\mu m^2 j_n(mx) [x j_n(x)]'- \mu_{sph} j_n(x) [mx j_n(mx)]'}{\mu m^2 j_n(mx) [x h_n^{(1)}(x)]'- \mu_{sph} h_n^{(1)}(x) [mx j_n(mx)]'} \label{eq_miea} \\
    b_n & = \dfrac{\mu_{sph} j_n(mx) [x j_n(x)]'- \mu j_n(x) [mx j_n(mx)]'}{\mu_{sph} j_n(mx) [x h_n^{(1)}(x)]'- \mu h_n^{(1)}(x) [mx j_n(mx)]'}, \label{eq_mieb}
\end{align}
\end{subequations}
where $m$ is the complex refractive index of the particle material, $\mu$ and $\mu_{sph}$ are the magnetic permeability of the medium and the particle, while $j_n(z)$ and $h_n^{(1)}(z)$ are $n$-th order Bessel and Hankel functions of the first kind, with the primed terms indicating derivatives with respect to the arguments. Using such coefficients, the scattering, extinction and absorption cross-sections can be expressed as:
\begin{subequations}
\begin{align}
    \sigma_{sca} & = \dfrac{2 \pi}{k^2} \sum_{n = 1}^\infty (2n + 1) ( |  a_n|^2 + |b_n|^2), \label{eq_mie_sca} \\
    \sigma_{ext} & = \dfrac{2 \pi}{k^2} \sum_{n = 1}^\infty (2n + 1) \mathfrak{Re} ( a_n + b_n), \label{eq_mie_ext} \\
    \sigma_{abs} & = \sigma_{ext} - \sigma_{sca}, \label{eq_mie_abs}
\end{align}
\end{subequations}
where we introduced the radiation wave number $k = 2 \pi / \lambda$. In principle, Eqs.~(\ref{eq_mie_sca}) and (\ref{eq_mie_ext}) imply the use of infinite series of summations. However, \citet{Wiscombe80} showed that the predicted results converge within the accuracy that can be represented by double precision floating point numbers by truncating the series at a maximum order $n = l_{max}$, given by:
\begin{equation}
    l_{max} = x + 4 x^{1/3} + 2. \label{eq_Wiscombe}
\end{equation}

While the Mie solution has very limited practical application, since it becomes quickly inaccurate as soon as the problem deviates from strict spherical symmetry, it nevertheless provides fundamental insight in the process of radiation extinction, because it can be used to define the scattering and absorption cross-sections of the \textit{equivalent sphere}, defined as a spherical particle that is made up by the same amount of material that composes a real particle, but has spherical shape. This concept acquires primary importance in the investigation of the role played by the particle structural features, since deviation from the spherical geometry may lead the same amount of material to be considerably more (or less) effective in interacting with radiation at different wavelengths, with respect to the ideal case, with obvious implications on the chemical properties inferred from the spectroscopic analysis of observations. For this reason, it is also customary to express the ability of particles to scatter, absorb and extinguish radiation in terms of \emph{efficiencies}, defined as the ratios between the scattering, the absorption, and the extinction cross-sections with respect to the particle's geometric cross-section $\sigma_g$:
\begin{fleqn}[\parindent]
\begin{equation}
\begin{split}
Q_{sca} = & \dfrac{\sigma_{sca}}{\sigma_g}; \\ 
Q_{abs} = & \dfrac{\sigma_{abs}}{\sigma_g}; \\ 
Q_{ext} = & Q_{sca} + Q_{abs} = \dfrac{\sigma_{ext}}{\sigma_g}.
\end{split}
\end{equation}
\end{fleqn}
These quantities can be used to derive another useful parameter, the so-called \textit{asymmetry parameter} $g = \langle \cos \theta \rangle$ \citep{Henyey41}, $\theta$ being the angle between incident and scattered radiation, which expresses the particle's average tendency to preferentially scatter in a forward direction ($g \rightarrow +1$), in a backward direction  ($g \rightarrow -1$), or rather isotropically  ($g \rightarrow 0$). In the case of a spherical particle, $g$ takes the form:
\begin{fleqn}[\parindent]
\begin{equation}
\begin{split}
    g = & \dfrac{4}{x^2 Q_{sca}} \left [ \sum_{n=1}^\infty \dfrac{n (n + 2)}{n + 1} \mathfrak{Re}\left(a_n a_{n + 1}^* + b_n b_{n + 1}^*\right) \right. \\ & \left. + \sum_{n=1}^\infty \dfrac{2n + 1}{n (n + 1)} \mathfrak{Re}\left(a_n b_n^*\right) \right], \label{eq_cosav}
\end{split}
\end{equation}
\end{fleqn}
where we used the $^*$ symbol to denote complex conjugation.

\subsection{Aggregate of spheres and the \textit{T-matrix} formalism}
\label{sec-tmatrix}

\begin{figure}
\begin{center}
\includegraphics[width=0.45\textwidth]{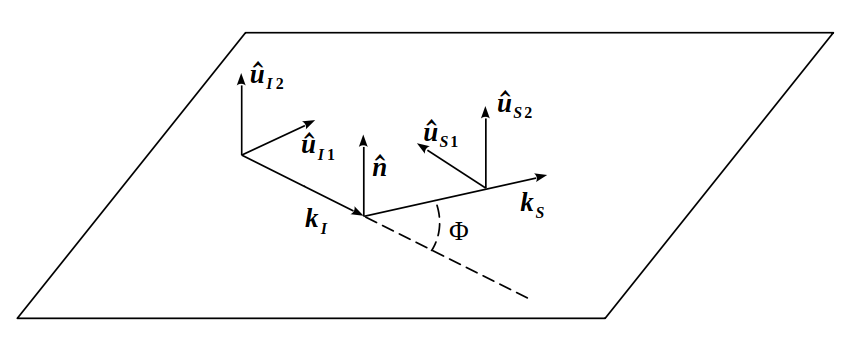}
\end{center}
\caption{The scattering plane. $\Phi$ is the scattering angle, $\bvec{k}_I$ is the incident wave vector, $\bvec{k}_S$ is the scattered wave vector, $\hat{\bvec{n}}$ is the plane normal unit vector, while $\hat{\bvec{u}}_{I \eta}$ and $\hat{\bvec{u}}_{S \eta}$ ($\eta = 1,2$) are the unit base vectors for the scattering process description. Adapted from \citet{Borghese07}. \label{fig_sca_plane}}
\end{figure}

If the particle is not spherically symmetric, the relation between the incident and the scattered fields cannot be derived analytically and the problem needs to be solved through numerical approaches. However, an analytical solution can be recovered if the particle is constituted (or can be approximated) by an aggregate of spherical units, which we refer to as \textit{monomers}. Following the approach described by \citet{Borghese79} and further detailed by \citet{Saija01}, it is possible to represent a particle of arbitrary shape as an aggregate of spherical monomers and to use the linearity of the field equations, which describe the field interaction with each monomer, to derive the total solution. Using the standard complex representation (where the physical fields are the real part of complex quantities), and assuming that the incident field $\bvec{E}_I$ is a polarized plane wave of the form:
\begin{equation}
    \bvec{E}_I = E_0 \hat{\bvec{e}}_I \exp(i \bvec{k}_I \cdot \bvec{r}) \label{eq_def_pol_field}
\end{equation}
(where we omit for convenience the $\exp(-i \omega t)$ factor giving the dependence on time), it is useful to express the polarization of the field with respect to the scattering plane, defined by the direction of incidence $\hat{\bvec{k}}_I$ and of scattering $\hat{\bvec{k}}_S$. We can then define two pairs of mutually orthogonal unit vectors $\hat{\bvec{u}}_{I,\eta}$ and $\hat{\bvec{u}}_{S,\eta}$ ($\eta = 1,2$), illustrated in Fig.~\ref{fig_sca_plane} and defined in such a way that $\hat{\bvec{u}}_{I,1}$ and $\hat{\bvec{u}}_{S,1}$ lie in the scattering plane, $\hat{\bvec{u}}_{I,2}$ and $\hat{\bvec{u}}_{S,2}$ are orthogonal to it, and:
\begin{subequations}
  \begin{align}
    \hat{\bvec{u}}_{I,1} \times \hat{\bvec{u}}_{I,2} & = \hat{\bvec{k}}_I \\
    \hat{\bvec{u}}_{S,1} \times \hat{\bvec{u}}_{S,2} & = \hat{\bvec{k}}_S.
  \end{align}
\end{subequations}
Then, the incident and the scattered fields $\bvec{E}_I$ and $\bvec{E}_{S}$ can be decomposed as:
\begin{subequations}
  \begin{align}
    \bvec{E}_I & = \sum_\eta (\bvec{E}_I \cdot \hat{\bvec{u}}_{I \eta}) \hat{\bvec{u}}_{I \eta} = \sum_\eta \bvec{E}_{I\eta} \\
    \bvec{E}_{S} & = \sum_{\eta} (\bvec{E}_{S} \cdot \hat{\bvec{u}}_{S\eta}) \hat{\bvec{u}}_{S\eta} = \sum_\eta \bvec{E}_{S\eta}
  \end{align}
\end{subequations}

Following \citet{Borghese07}, we can expand the incident electric field as:
\begin{equation}
  \bvec{E}_{I\eta}(\bvec{r}) = E_{0\eta} \sum_{plm} \bvec{J}^{(p)}_{lm}(\bvec{r}, k) W^{(p)}_{I\eta lm}, \label{eq_inc_exp}
\end{equation}
where $\bvec{J}^{(p)}_{nlm}$ are vector multipole fields, while the amplitudes of the incident field are $W^{(p)}_{I\eta lm} = W^{(p)}_{lm}(\hat{\bvec{u}}_{I\eta}, \hat{\bvec{k}}_I)$, with:
\begin{equation}
  W^{(p)}_{lm}(\hat{\bvec{e}}, \hat{\bvec{k}}) = 4\pi i^{p + l - 1} \hat{\bvec{e}} \cdot \bvec{Z}^{(p)*}_{lm}(\hat{\bvec{k}}). \label{eq_transverse}
\end{equation}
These amplitudes are such that the result of the linear combination of $\bvec{J}_{lm}^{(p)}$ in Eq.~(\ref{eq_inc_exp}) is aligned with $\hat{\bvec{u}}_{I\eta}$. In Eq.~(\ref{eq_transverse}), $\bvec{Z}^{(p)*}_{lm}(\hat{\bvec{k}})$ are transverse harmonics \citep{Fucile97, Saija03b}. Similarly, the scattered field can be expressed as:
\begin{equation}
  \bvec{E}_{S\eta}(\bvec{r}) = E_{0\eta} \sum_{plm} \bvec{H}^{(p)}_{lm}(\bvec{r}, k) A^{(p)}_{\eta lm},
\end{equation}
where $A^{(p)}_{\eta lm}$ are the amplitudes of the scattered field. Thanks to the linearity of the Maxwell equations and enforcing continuity on the surface of the particle, such amplitudes can be expressed as a function of the incident field's amplitudes introducing the T-matrix \citep{Waterman71}:
\begin{equation}
  A^{(p)}_{\eta lm} = S^{(pp')}_{lml'm'} W^{(p')}_{I\eta l' m'}. \label{eq_intro_tm}
\end{equation}
The elements of the T-matrix, $S^{(pp')}_{lml'm'}$, contain all the information concerning the particle morphology and its orientation with respect to the incident radiation field. The existence of such a T-matrix linear operator connecting the incident and scattering amplitudes implies no assumption, neither on the particle nor on the incident field, as it just stems from the linearity of Maxwell's equations. For spherical particles, the T-matrix approach exactly reproduces the Mie solution of the light scattering problem. 

Let's assume that the particle interacting with the incident wave of Eq.~(\ref{eq_def_pol_field}) can be modeled as an aggregate of spheres, each identified by an index $\alpha$, having radius $\rho_\alpha$, center coordinates $\bvec{R}_\alpha$, refractive index $n_\alpha$, and all embedded in a non-absorbing medium with refractive index $n$.  The incident field can still be expanded as in Eq.~(\ref{eq_inc_exp}), while the scattered field takes the form:
\begin{equation}
  \bvec{E}_{S\eta} = E_{0 \eta} \sum_\alpha \sum_{plm} \bvec{H}^{(p)}_{lm}(\bvec{r}_\alpha, k) \mathcal{A}^{(p)}_{\eta\alpha lm}, \label{eq_sca_exp}
\end{equation}
where $\bvec{r}_\alpha = \bvec{r} - \bvec{R}_\alpha$. On the other hand, the field inside the $\alpha^\mathrm{th}$ sphere expands to:
\begin{equation}
  \bvec{E}_{T\eta\alpha} = E_{0 \eta} \sum_\alpha \sum_{plm} \bvec{J}^{(p)}_{lm}(\bvec{r}_\alpha, k_\alpha) \mathcal{C}^{(p)}_{\eta\alpha lm}, \label{eq_inter_exp}
\end{equation}
with $k_\alpha = n_\alpha k$. The amplitudes $\mathcal{A}^{(p)}_{\eta\alpha lm}$ and $\mathcal{C}^{(p)}_{\eta\alpha lm}$ in Eqs.(\ref{eq_sca_exp}) and (\ref{eq_inter_exp}) are determined by the boundary conditions at the surface of each sphere and their knowledge solves the problem of scattering by the whole aggregate. In fact, the amplitudes $\mathcal{A}^{(p)}_{\eta\alpha lm}$ are the solution of the linear system of non-homogeneous equations \citep{Borghese84}:
\begin{equation}
  \sum_{\alpha'} \sum_{p' l' m'} \mathcal{M}^{(pp')}_{\alpha l m \alpha' l' m'} \mathcal{A}^{(p')}_{\eta \alpha' l' m'} = -\mathcal{W}^{(p)}_{\eta \alpha l m}, \label{eq_lin_system}
\end{equation}
where we have introduced the \textit{shifted} amplitudes of the incident field:
\begin{equation}
  \mathcal{W}^{(p)}_{\eta \alpha l m} = \sum_{p' l' m'} \mathcal{J}^{(pp')}_{\alpha l m 0 l' m'} W^{(p')}_{I\eta l' m'}, \label{eq_shifted_ampl}
\end{equation}
and:
\begin{equation}
  \mathcal{M}^{(p p')}_{\alpha l m \alpha' l' m'} = \left[ R^{(p)}_{\alpha l} \right]^{-1} \delta_{\alpha \alpha'} \delta_{p p'} \delta_{l l'} \delta_{m m'} + \mathcal{H}^{(p p')}_{\alpha l m \alpha' l' m'}. \label{eq_deltas}
\end{equation}
The quantities $\mathcal{J}^{(pp')}_{\alpha l m 0 l' m'}$ are the elements of the matrix that translates the multipole fields from the origin of the coordinates to the center of the $\alpha^\mathrm{th}$ sphere $\bvec{R}_\alpha$, according to the multipole field addition theorem \citep{Borghese80}. The quantities $R^{(1)}_{\alpha l}$ and $R^{(2)}_{\alpha l}$, except for a sign, are the elements of the T-matrix of the $\alpha^\mathrm{th}$ sphere, while the quantities $\mathcal{H}^{(p p')}_{\alpha l m \alpha' l' m'}$ are also derived from the vector multipole field addition theorem and take into account the effects of multiple scattering processes between the spheres in the aggregate \citep{Borghese94}.

In order to calculate the T-matrix of the whole aggregate, we define $\bmat{M}$ as the matrix of the coefficients introduced by Eq.~(\ref{eq_deltas}). Then, the formal analytical solution of Eq.~(\ref{eq_lin_system}) is:
\begin{equation}
  \mathcal{A}^{(p)}_{\eta \alpha l m} = -\sum_{p' l' m'} \left[ \bmat{M}^{-1} \right]^{(p p')}_{\alpha l m \alpha' l' m'} \mathcal{W}^{(p')}_{\eta \alpha' l' m'}. \label{eq_form_sol}
\end{equation}
The addition theorem for vector multipole fields also allows us to write the scattered field of Eq.~(\ref{eq_sca_exp}) in terms of multipole fields with origin coincident with the origin of the coordinate system $O$:
\begin{equation}
  \bvec{E}_{S\eta} = E_{0 \eta} \sum_{p l m} \sum_{\alpha'} \sum_{p' l' m'} \bvec{H}^{(p)}_{lm}(\bvec{r}, k) \mathcal{J}^{(p p')}_{0 l m \alpha' l' m'} \mathcal{A}^{(p')}_{\eta \alpha' l' m'},
\end{equation}
although this solution only applies out of the smallest sphere that contains the whole particle (i.e., it is appropriate to calculate the scattered field in the far zone). As a consequence the amplitudes of the field scattered by the whole aggregate take the form:
\begin{equation}
  A^{(p)}_{\eta' l m} = \sum_{\alpha'} \sum_{p' l' m'} \mathcal{J}^{(p p')}_{0 l m \alpha' l' m'} \mathcal{A}^{(p')}_{\eta' \alpha' l' m'}, \label{eq_alt_ampl}
\end{equation}
where, this time, the quantities $\mathcal{J}^{(p p')}_{0 l m \alpha' l' m'}$ are the elements of the matrix that transfers the origin of the $\bvec{H}$-multipole fields from $\bvec{R}_{\alpha'}$ back to the origin of the coordinates. At this point, comparing Eq.~(\ref{eq_intro_tm}) with Eq.~(\ref{eq_alt_ampl}), using Eq.~(\ref{eq_form_sol}) to express $\mathcal{A}^{(p')}_{\eta' \alpha' l' m'}$ and Eq.~(\ref{eq_shifted_ampl}) to express $\mathcal{W}^{(p)}_{\eta \alpha l m}$, we obtain the T-matrix elements of the whole aggregate as:
\begin{equation}
  S^{(p p')}_{l m l' m'} = \sum_{\alpha \alpha'} \sum_{q L M} \sum_{q' L' M'} \mathcal{J}^{(p q)}_{0 l m \alpha L M} \left[ \bmat{M}^{-1} \right]^{q q'}_{\alpha L M \alpha' L' M'} \mathcal{J}^{(q' p')}_{\alpha' L' M' 0 l' m'}. \label{eq_tm_def}
\end{equation}

The T-matrix elements depend on the reference frame used to describe the scattering process. However, since the vector spherical multipoles are eigenfunctions of the rotation matrices, their orientational averages can be obtained analytically. Knowledge of the T-matrix allows for the calculation of the scattering amplitude for any given combination of particle geometry and incident and scattered radiation directions. These amplitudes eventually lead to the formulation of the scattering and extinction cross-sections, associated with any given polarization state, as \citep{Borghese07}:
\begin{subequations}
    \begin{align}
        \sigma_{ext|\eta \eta'} & = - \dfrac{1}{k^2} \sum_{plm} \sum_{p' l' m'} W_{I \eta l m}^{(p)*} \mathcal{S}_{l m l' m'}^{(pp')} W_{I \eta' l' m'}^{(p')} \\
        \sigma_{sca|\eta \eta'} & = \dfrac{1}{k^2} \sum_{plm} \sum_{p' l' m'} \sum_{p'' l'' m''} \mathcal{S}_{l m l' m'}^{(pp')} W_{I \eta l' m'}^{(p')*} \mathcal{S}_{l m l'' m''}^{(pp'')} W_{I \eta' l'' m''}^{(p'')},
    \end{align}
\end{subequations}
with $\sigma_{abs|\eta \eta'} = \sigma_{ext|\eta \eta'} - \sigma_{sca|\eta \eta'}$, as before.

\begin{figure*}[t]
\begin{center}
\includegraphics[width=\textwidth]{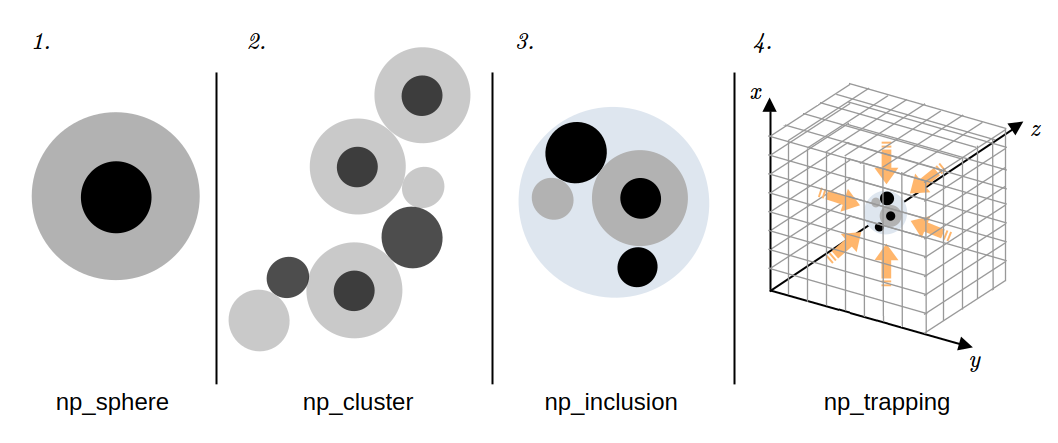}
\caption{Schematic representation of the modeling possibilities of the \nptm\ application suite. The case of a single particle with spherical symmetry and one or more concentric layers is addressed by \texttt{np\_sphere} in the framework of the Mie theory (panel $1$). \texttt{np\_cluster} solves the case of aggregates of arbitrary numbers of spherical monomers, with the possibility of using monomers with different compositions, concentric layer structures and sizes (panel $2$). \texttt{np\_inclusion} is similar to \texttt{np\_cluster} except for embedding the aggregate in an external spherical coating (panel $3$). Finally, \texttt{np\_trapping} solves for the forces and torques exerted on a particle trapped by a radiation beam (panel 4). \label{fig_np_apps}}
\end{center}
\end{figure*}
\subsection{Radiation forces}
\label{sec_radforces}

The processes of scattering, absorption and emission of radiation by material particles affect their dynamical and thermal properties. In particular, the effect of extinction depends critically on the cross-section of particles, as seen by the incident radiation field. The interaction of a radiation field with intensity $I_0$, propagating along a direction of incidence defined by the unit vector $\hat{\bvec{k}}_I$ on a particle leads to a net mechanical force \citep{Mishchenko01, Polimeno18}:
\begin{equation}
  \bvec{F}_I = \frac{I_0}{c} \left[\sigma_{ext} \hat{\bvec{k}}_I - \int_\Omega \left(\frac{\de \sigma_{sca}}{\de \Omega} \right) \hat{\bvec{k}}_S \de \Omega \right],
\end{equation}
where $\hat{\bvec{k}}_S$ is the unit vector pointing in the direction, as given by the integration variable $\Omega$, of scattered light.
In general, $\bvec{F}_I$ is not oriented along the same direction of incidence $\hat{\bvec{k}}_I$, as it depends also on the scattering efficiency. Its component along the $\hat{\bvec{k}}_I$ direction is the radiation pressure force. For a scattering angle $\Phi$, such component can be expressed as:
\begin{fleqn}[\parindent]
\begin{equation}
\begin{split}
  F_{pr} = \bvec{F} \cdot \hat{\bvec{k}}_I & = \frac{I_0}{c} \left[ \sigma_{ext} - \int_\Omega \cos \Phi \left(\frac{\de \sigma_{sca}}{\de \Omega} \right) \de \Omega \right] \\ & = \frac{I_0}{c} [ \sigma_{ext} - g \sigma_{sca} ], \label{eq_rad_pres}
\end{split}
\end{equation}
\end{fleqn}
where $g$ is the asymmetry parameter for scattering at an angle $\Phi$:
\begin{equation}
  g = \frac{1}{\sigma_{sca}} \int_\Omega \cos \Phi \left( \frac{\de \sigma_{sca}}{\de \Omega} \right) \de \Omega.
\end{equation}

In a completely analogous way, using Kirchoff's law and the detailed balance principle, one can derive an expression for the net force due to thermal emission by the particle. Thermal emission is given by
\begin{equation}
  I_{th}(\nu, \Omega; T) = I_\mathrm{bb}(\nu; T) \left(\frac{\de \sigma_{abs}}{\de \Omega} \right),
\end{equation}
where $I_\mathrm{bb}(\nu; T)$ is Planck's blackbody spectrum at temperature $T$ and frequency $\nu$. This emission can be anisotropic, and thus result in a net recoil force given by
\begin{equation}
  \bvec{F}_\mathrm{th} = - \left\{\int_{\nu=0}^{\infty} \frac{I_\mathrm{bb}(\nu; T)}{c} \left[ \int_\Omega \left(\frac{\de \sigma_{abs}}{\de \Omega} \right) \hat{\bvec{k}}_\Omega \de \Omega \right] \de \nu \right\},
\end{equation}
where $\hat{\bvec{k}}_\Omega$ is the unit vector in the direction of the integration solid angle element $\Omega$.

If the distribution of particle geometry and orientations can be represented in terms of analytical models, the T-matrix formalism allows for the calculation of the average expected cross-sections, leading to re-write the net radiation pressure effect of Eq.~(25) as: 
\begin{equation}
  \langle F_{pr} \rangle = \frac{I_0}{c} [ \langle \sigma_{ext} \rangle - \langle g \rangle \langle \sigma_{sca} \rangle ], \label{eq_pres_mean}
\end{equation}
where the terms in brackets represent direction averaged quantities \citep{Polimeno21}.

One of the most remarkable advantages of the T-matrix formalism is that all the solicitations that take place can be evaluated across the whole structure of the particle. This, in turn, provides the opportunity to derive the radiative stresses and torques that affect the modeled structure and include them in the results.

\section{Code implementation}
\label{sec-code}

\subsection{Project structure and goals}
The \nptm\ project\footnote{\nptm\ is an OpenSource project distributed under GNU GPLv3 license via \texttt{gitLab} and \texttt{gitHub} respectively at:\\ \url{https://www.ict.inaf.it/gitlab/giacomo.mulas/np_tmcode}\\ \url{https://github.com/ICSC-Spoke3/NP_TMcode}} aims at reimplementing the T-matrix formalism for the calculation of extinction and radiation forces and torques, rewriting the code originally developed by \citet{Saija01} and \citet{Borghese07} in \texttt{FORTRAN~66}, in a parallel and scalable version, with the goal of optimizing the use of hardware resources and reducing the computing time required to solve particle models. The code is subdivided in a suite of applications and computing libraries, written in \texttt{C++}, with the addition of a set of auxiliary \texttt{python3} scripts, used to assist users in the definition of the input data and in the inspection of the results. In addition, the process of porting the code to a modern language was carried out together with the construction of a complete code inline documentation, thus producing a more easily maintainable solution. The application suite includes four programs: 
\begin{itemize} 
\item \texttt{np\_sphere}, for the solution of the case of a single particle with spherical symmetry (according to the formalism laid out in section~\ref{sec-mie}),
\item \texttt{np\_cluster}, which solves the case of a generic particle represented by an arbitrary combination of spherical components, using the formalism presented in sec.~\ref{sec-tmatrix},
\item \texttt{np\_inclusion}, formally similar to \texttt{np\_cluster}, but adding a spherical outer coating around the aggregate,
\item \texttt{np\_trapping}, which, instead, calculates the forces and the torques exerted on a particle trapped by a focused laser beam with the ``optical tweezers'' technique. 
\end{itemize}

For convenience, these four core applications follow the same execution scheme, which uses a set of machine readable configuration files, to describe the input model, and then writes the results in an output folder. The output consists of the integrated and differential cross-sections, the asymmetry parameters and the forces and torques exerted on a model particle exposed to a radiation field. A scheme of the situations addressed by the \nptm\ applications is shown in Fig.~\ref{fig_np_apps}.

\subsection{Numeric optimization}

The solution of the scattering process and its application to experimental and observational scenarios involves a sequence of well defined steps, which can be identified as the imposition of the boundary conditions that enforce field continuity across the particle and for a physical solution at large distances, eventually producing the $\bmat{M}$ matrix of Eq.~(\ref{eq_form_sol}), the inversion of $\bmat{M}$ to obtain the T-matrix of Eq.~(\ref{eq_tm_def}), and, finally, its use in the derivation of integrated and possibly differential cross-sections. In the vast majority of cases, the most challenging computational task is represented by the matrix inversion stage, unless a very large number of differential quantities is requested by the user. This stage was therefore the target of the first optimization efforts, which were pursued by combining specifically optimized algebraic libraries, such as \lapack\footnote{see \url{https://www.netlib.org/lapack/} for the definition and a reference implementation, and \url{http://www.openmathlib.org/OpenBLAS/} or \url{https://www.intel.com/content/www/us/en/developer/tools/oneapi/onemkl.html} for some optimised implementations} \citep{lapack99} or \magma\footnote{see \url{https://icl.utk.edu/}magma/} \citep[for systems featuring a suitable GPU architecture,][]{magma97},\footnote{\nptm\ applications can use different \lapack\ and \magma\ functions to invert a complex matrix (depending on the availability of these packages and on the presence of a GPU). If none of these is available, an internal $LU$ factorization fallback algorithm is used.} with parallel execution strategies.

\begin{figure}[t]
\begin{center}
\includegraphics[width=0.445\textwidth]{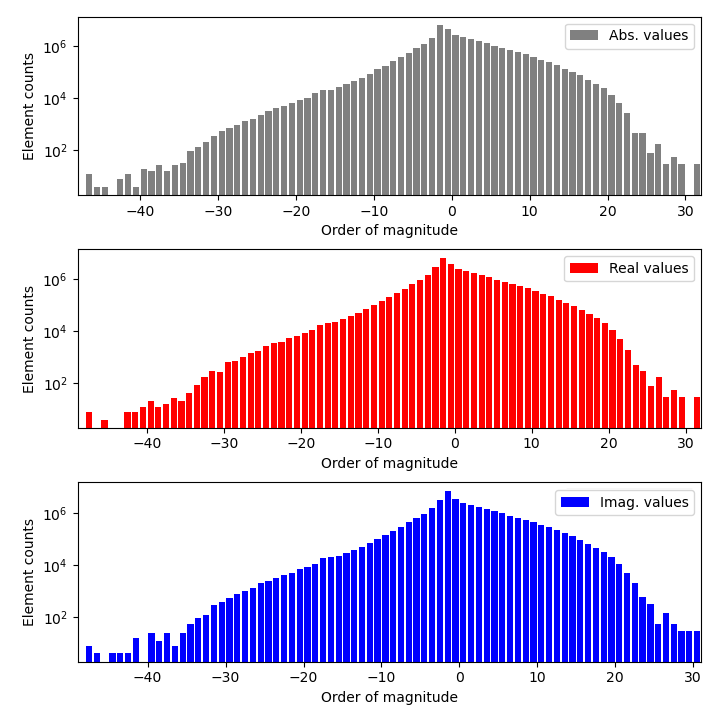}
\caption{Histograms of the orders of magnitude of the complex coefficients for the T-matrix of a model particle made up by $13$ spherical monomers, computed with $l_{max} = 14$. The magnitudes of the absolute values are shown in the top panel, those of the real parts are given in the middle panel, while those for the imaginary parts are shown in the bottom panel. \label{fig_dyn_range}}
\end{center}
\end{figure}
Speaking in numeric terms, the T-matrix is a dense complex matrix, whose coefficients depend on the properties of the model particle, represented as an aggregate of spheres with known positions, sizes, layering structures and refractive indices. The dimensions $[N \times N]$ of the T-matrix are controlled by the number of spherical monomers $n_{sph}$, used to model the particle, and by the multipole field expansion order truncation $l_{max}$, resulting in
\begin{equation}
N = 2 n_{sph} l_{max} (l_{max} + 2)
\end{equation}
distinct elements in Eq.~(\ref{eq_tm_def}) \citep{Borghese07}.

\begin{figure*}[t]
\begin{center}
\includegraphics[width=0.95\textwidth]{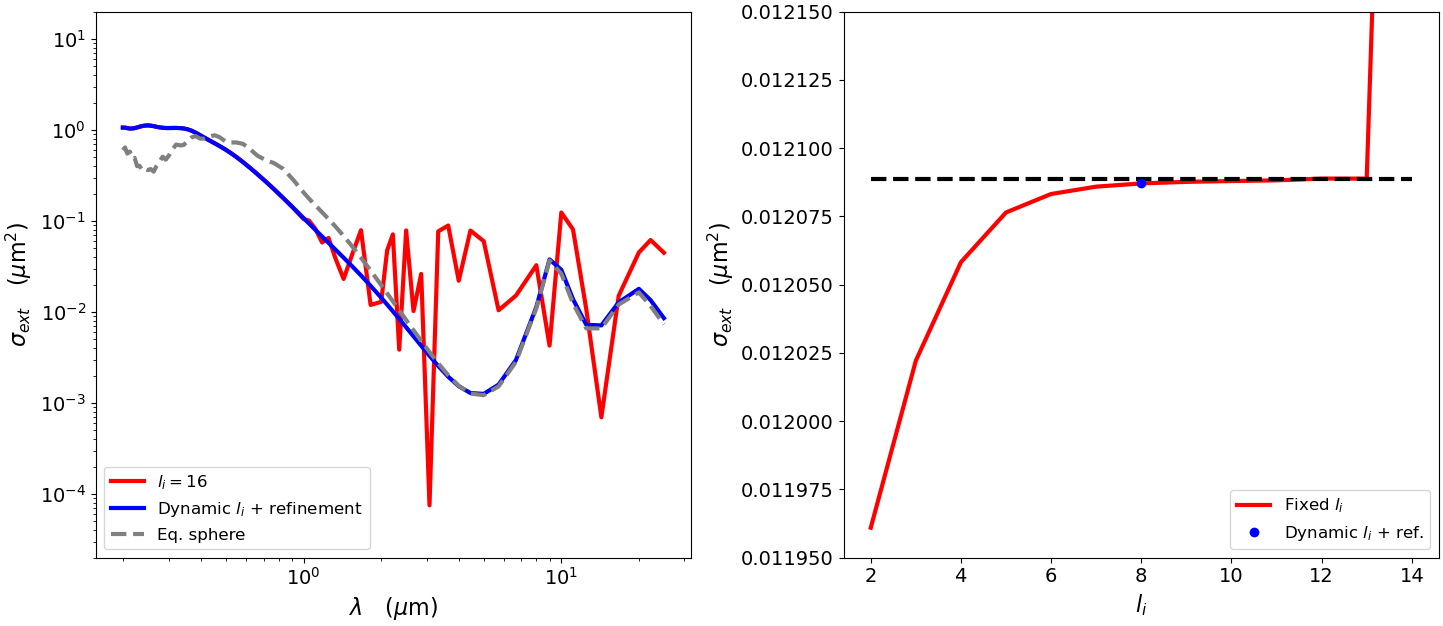}
\caption{Effect of numeric instability on the calculation of $\sigma_{ext}$ for an aggregate of $16$ enstatite spheres ($r_{sph} = 0.1\, \mu$m). Left panel: calculation over the wavelength range $0.188\, \mu\mathrm{m} \leq \lambda \leq 25.0\, \mu\mathrm{m}$. Using a fixed expansion order results in instability at long wavelength (red line), while dynamic order assignment combined with iterative refinement recovers a stable solution (blue line). The gray dashed line is the cross-section of an equivalent mass enstatite sphere with $r_{sph} = 0.25\, \mu$m. Right panel: extinction cross-section $\sigma_{ext}$ at $\lambda = 2.11\, \mu\mathrm{m}$ computed in dynamic order (blue point) and for different values of fixed $l_i$ (red curve). The black dashed line represents the convergence limit. In this case, instability arises with $l_i > 13$. \label{fig_instability}}
\end{center}
\end{figure*}
For a single sphere, $l_{max}$ can be directly derived from Eq.~(\ref{eq_Wiscombe}). However, in the case of an aggregate of spheres, where the radius used to define the size parameter in Eq.~(\ref{eq_size_par}) becomes the radius of the smallest sphere encompassing the particle, the necessary order of truncation grows substantially, especially for wavelength regimes which are comparable to the particle size or shorter. The subsequent dramatic increase of $N$ poses two fundamental problems. The first obvious issue is the total size of the T-matrix, which, if represented with double precision complex elements, can easily require several gigabytes of memory to be handled, even for models using only few tens of spherical monomers. The second problem, instead, stems from the large dynamic range of its elements, whose harmonic nature implies that, for large values of $l_{max}$, coefficients separated by more and more orders of magnitude need to be accounted for, as it is illustrated for example in Fig.~\ref{fig_dyn_range}. Indeed, since the physically meaningful quantity, namely the T-matrix, is the inverse of the matrix obtained by enforcing continuity conditions for the fields, adding smaller and smaller corrections to the T-matrix with increasing order of the expansions results in \emph{larger} terms being added to the matrix \emph{before} inversion. Given the limited dynamic range represented by double precision floating point variables, the loss of accuracy in the description of the matrix coefficients results in the matrix inversion becoming numerically unstable. This means that one has to increase the truncation order until the truncation error becomes low enough for the desired accuracy, but stop before numerical instability sets in. The larger the monomer spheres, the larger the order needs to be before convergence is achieved (see Eq.~\ref{eq_Wiscombe}). An example of the effects of numerical instability is shown in Fig.~\ref{fig_instability} for a small cluster of 16 enstatite spheres. In particular, the right panel shows that for this specific model, at $\lambda = 2.11\, \mathrm{\mu m}$, numerical convergence is achieved at $l_i = 8$, in agreement with Wiscombe's criterion (Eq.~\ref{eq_Wiscombe}), while numerical instability sets in at $l_i > 13$, giving a "good" interval of 6 orders yielding accurate results.

In all cases of low conductivity materials, numerical instabilities appear at truncation orders much higher than the smallest one required for convergence.\footnote{High conductivity materials, where Wiscombe's criterion does not equally apply, tend to become unstable much closer to the convergence value. This issue is further addressed in Appendix A.} \nptm\ deals with numeric instability through the combination of two optional techniques, namely the dynamic adjustment of $l_{max}$, which is then determined independently on the fly for each wavelength, and matrix inversion iterative refinement, which, on the contrary, expands the range of stable orders, if the model is formed by monomers with remarkably different sizes. The $l_{max}$ dynamic computation is based on the approach laid down by \citet{Saija03a} and \citet{Iati04}, who demonstrated that the matrix inversion stage does not need to account for the whole particle, but it can be applied while still considering the particle spherical components. This allows for the use of a smaller internal maximum multipole field expansion order, named $l_i$, and then to subsequently combine the results in a matrix using a separate external truncation order $l_e$, chosen to adequately represent the particle as a whole. Knowledge of these fundamental parameters allows for the solution of the problem for each relevant radiation wavelength.

If, however, the particle model is composed by monomers with very different size, which may require different orders to be solved, iterative matrix inversion refinement can be used to improve the stability of the calculation. Provided that the first matrix inversion $\hat{\bmat{A}}_k$ is a reasonable numeric approximation of the real inverse matrix $\bmat{A}^{-1}$, we can define a residual matrix $\bmat{R}_k$:
\begin{equation}
    \bmat{R}_k = I - \bmat{A} \cdot \hat{\bmat{A}}_k, \label{eq_mat_residual}
\end{equation}
where $I$ is the identity matrix, such that the new matrix:
\begin{equation}
    \hat{\bmat{A}}_{k + 1} = \hat{\bmat{A}}_k \cdot \bmat{R}_k + \hat{\bmat{A}}_k \label{eq_mat_refine}
\end{equation}
will be a more accurate approximation of the inverse. Choosing one of the many available numeric implementations of complex matrix inversion to obtain a first numeric approximation of the inverted matrix in Eq.~(\ref{eq_tm_def}), \nptm\ applications can iterate Eqs.~(\ref{eq_mat_residual}) and (\ref{eq_mat_refine}) on $k$ until either $\bmat{R}$ contains only elements whose absolute value lies below a predefined threshold, or a maximum number of refinement iterations was reached (in which case a warning message is issued).

\subsection{Parallelization strategy}

The \nptm\ applications were designed to be optimized and scalable, in order to improve over the performance of the original algorithms and to take full advantage of parallel hardware architectures. Depending on whether the execution is planned on single workstations or on multiple nodes of a computing farm, multiple layers of parallelization were introduced. A configuration script scans the hardware capabilities of the host system and looks for the software needed to enable optional optimization features, in order to assist users in the code set up and installation.

In most cases, the scientific application motivating the analysis involves the evaluation of the particle scattering and absorption properties when interacting with radiation fields of different wavelengths. Once the particle model is fixed and the optical properties of the composing materials are known for each wavelength in the calculation, this is an embarrassingly parallel problem, which can be effectively accelerated by using multiple threads on a workstation (or a computing node) and multiple processes spread across different nodes.\footnote{To be more precise, multi-process wavelength parallelism on a single computing system is also supported, as a consequence of the code scalability.} \texttt{np\_sphere}, \texttt{np\_cluster}, and \texttt{np\_inclusion} take advantage from this fact by encapsulating the data that are needed to solve the problem for every wavelength in classes that expose methods to be distributed across multiple threads via \omp\footnote{see \url{https://www.openmp.org/}} directives \citep{Dagum98} and multiple processes through \mpi\footnote{see \url{https://www.mpi-forum.org/}} functions \citep{MPI93}. Additional levels of multi-threaded parallelism are managed via \omp\ hierarchical multi-threading or via GPU offload of \textit{single instruction on multiple data} (SIMD) loops and of algebraic operations involving large vectors and matrices.

A considerably different strategy is adopted in the case of \texttt{np\_trapping}. In this case, the application works with the precomputed scattering properties of a particle model at a specific wavelength, to determine the dynamic effects of its interaction with a corresponding radiation beam. As illustrated in the rightmost panel of Fig.~\ref{fig_np_apps}, the solution of the problem is derived by computing the characteristics of the radiation field on a grid of points, defined by the user, which encompasses the region where the model particle is placed.  By convention, the code defines the primary beam propagation direction as the $z$-axis of the modeled space. In this case, the most computationally demanding task is the evaluation of the field effects in the different grid points. This also represents an embarrassingly parallel problem, which lies at the tip of the parallel execution hierarchy of \texttt{np\_trapping}, and is handled via \omp\ directives designed to work optimally either with multi-core CPUs or with GPUs. Two additional parallel layers are used to operate on large vectors and to perform algebraic calculations with vectors and matrices. These levels are explicitly handled via \omp\ directives on CPUs, after introducing a SIMD code format for the relevant loops, while they are implicitly organized on a best effort basis via an \texttt{omp target teams distribute parallel for} directive on a linearized loop structure on GPUs. Once the field characteristics are known, the user can apply different model particles, eventually obtaining the forces and torques that these samples would experience in the various positions defined by the grid.

\begin{table*}[t]
\begin{center}
\caption{Hardware configurations used for code testing and performance benchmarking. The table columns report, respectively, (i) a configuration identifier, (ii) the available CPUs, (iii) the total number of CPU cores (physical/logical), (iv) the host system RAM, (v) the number of available GPU devices, (vi) the GPU RAM on each GPU device, and (vii) the GPU model. \label{tab-hardware}}
\begin{tabular}{lcccccc}
    \hline
    \hline
    ID & CPU & CPU cores & Host RAM & No. of GPUs & GPU RAM & GPU model \\
    \hline
    A & Intel Core i9-13900H & 14/20$^{(1)}$ & 32 Gb & 1 & 8 Gb & \begin{tabular}{@{}c@{}}NVIDIA GeForce RTX 4060 \\ (Ada Lovelace)\end{tabular} \\
    B$^{(2)}$ & 2 $\times$ AMD Epyc 7313 & 32/64$^{(3)}$ & 503 Gb & 2 & 96 Gb$^{(4)}$ & NVIDIA A40 (Amp\`ere) \\
    \hline
\end{tabular}
\end{center}
\begin{footnotesize}
$^{(1)}$ Intel Core 19-13900H processors have 6 hyper-threading cores and 8 single threading cores. \\
$^{(2)}$ The system description parameters refer to the hardware of a single computing node in the cluster. \\
$^{(3)}$ AMD Epyc 7313 processors support Simultaneous Multi-Threading (SMT), allowing each physical CPU core to execute 2 simultaneous threads. \\
$^{(4)}$ 48 Gb of RAM per GPU.
\end{footnotesize}
\end{table*}
\section{Performance analysis}
\label{sec-perf}

Porting the original \texttt{FORTRAN~66} applications to a parallel \texttt{C++} implementation required the refactoring of some algorithms and the migration of the data structures towards a reorganized class hierarchy. While this effort granted the possibility to improve the code functionality with the addition of runtime configuration options, it introduced slight overheads, of the order of few seconds, in the execution of the serial steps of the \texttt{C++} implementation, with respect to the corresponding \texttt{FORTRAN~66} instructions. Such very minor overheads, however, are vastly compensated by the reduction of the user workload in the input preparation stage and in the management of the final output. In addition, the considerable performance gain and scalability offered by the activation of the code parallel sections results in a substantial improvement of the domain of practical applicability of the T-matrix formalism.

The main parameters that we can consider to quantify the performance gain of \nptm\ over previous implementations of the T-matrix formalism include the comparison of the execution time required to solve preexisting development test models, the assessment of the result accuracy and the ability of the code to adapt to scalable hardware resources. In its most direct set up, \nptm\ is designed to run under \textit{UNIX}-like operating systems, although configurations that use container versions handled through \textit{Docker}\footnote{\url{https://www.docker.com/}} and \textit{Singularity}\footnote{\url{https://sylabs.io/}} allow for execution on a wider range of system architectures. In this section, however, we focus our attention on the performance of the \nptm\ applications in their native configuration, comparing the execution time and the results obtained on the same hardware, when running development test cases with the serial and the parallel versions. Then we assess the performance scalability, by using different hardware resources to solve more advanced particle models, for which complete solutions via the serial version cannot be obtained in reasonable execution times. The test cases that we directly include in this work are solved by \texttt{np\_cluster} and \texttt{np\_trapping} and subsequently compared with the original code executed on the same hardware.\footnote{The case of \texttt{np\_sphere} is not directly considered here because the machine implementation of the Mie formalism of Sec.~\ref{sec-mie} is extremely fast and requires only fractions of second to be executed. \texttt{np\_inclusion}, on the other hand, is formally similar to \texttt{np\_cluster}.} Depending on the complexity of the models adopted to benchmark the code, we ran the calculations either on a commercial laptop, namely an ASUS ZenBook Pro (hereafter hardware configuration A, mostly used for development cases), or using one or more nodes of the computing farm hosted at INAF - Astronomical Observatory of Cagliari (hereafter hardware configuration B, used for benchmarking and scaling tests of more advanced cases). The main hardware features of these configurations are listed in Table~\ref{tab-hardware}. We here show only a subset of the many test cases we ran, omitting the very short ones we internally used only for basic consistency tests. 

\begin{figure}[t]
\begin{center}
\includegraphics[width=0.45\textwidth]{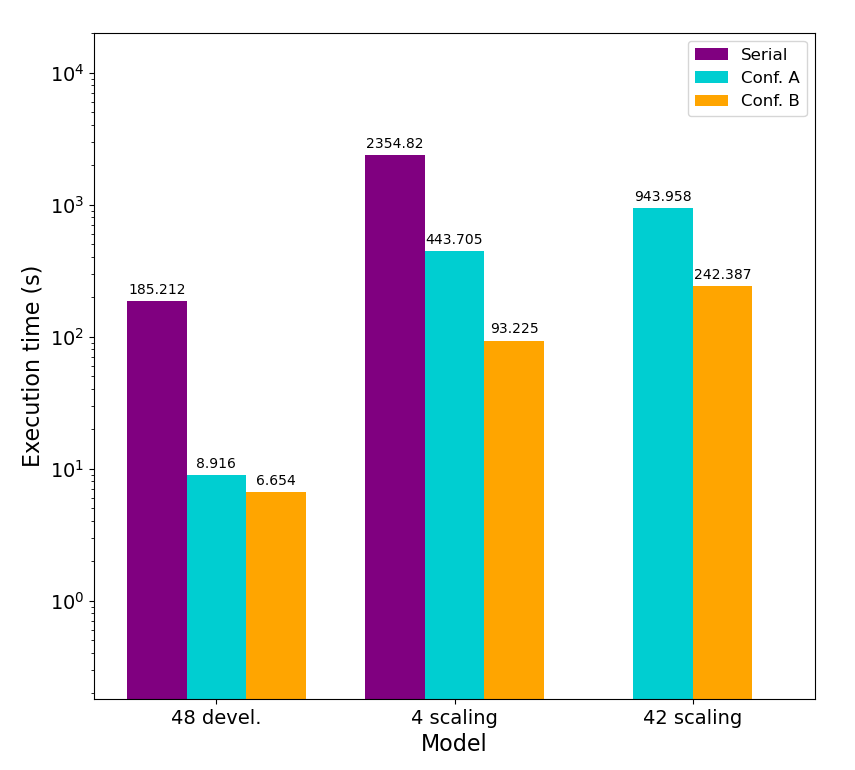}
\caption{Comparison of aggregate particle model calculation times, illustrating the performance of the original serial code with respect to \texttt{np\_cluster} parallel implementations, executed with hardware configurations A and B (the $y$-axis is plotted in logarithmic scale). A detailed description of the fundamental characteristics of each model is given in Table~\ref{tab_models}. \label{fig_np_cluster_bench}}
\end{center}
\end{figure}
\subsection{\texttt{np\_cluster} benchmarks}

In order to thoroughly and effectively test the software performance we need a set of input models, which should be ideally quick to run (in order to be compared with well tested serial implementations) and explore the largest possible extent of the modeling parameter space. Such \textit{development models} are generally limited to small particles, having a low number of constituents, but with a large variety of sphere types, with different compositions, sizes and layering structures. These models are typically evaluated on small, but critical ranges of the electromagnetic radiation frequency spectrum, where substantial changes in the material optical properties are expected, and they can be handled by hardware configuration A. To further assess the code scalability properties, we also use more advanced models, typically made up by a large number of spherical constituents, but with fewer types of components. These \textit{scaling models} are generally (but not exclusively) executed in hardware configuration B. The performance of the serial implementation was evaluated running the original codes as single-threaded applications on a single CPU core of configuration A, for development models, and a single CPU core of configuration B, for scaling models. However, not all the scaling models could be solved with the serial implementation, because some would just require an unreasonably long time to complete.

\begin{figure}[t]
\begin{center}
\includegraphics[width=0.45\textwidth]{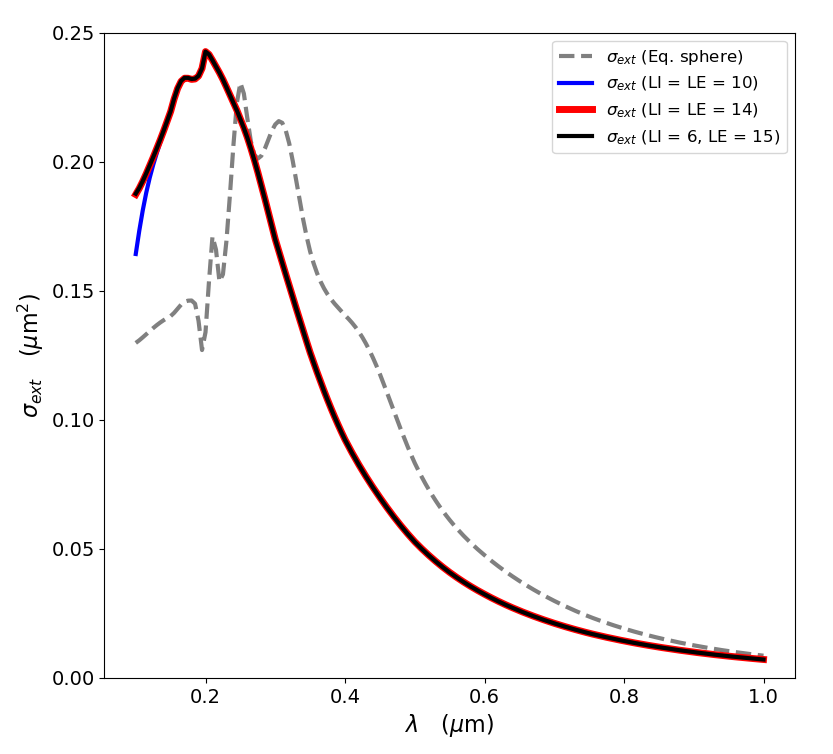}
\caption{The extinction cross-section $\sigma_{ext}$ as a function of wavelength $\lambda$ for the particle model named \textit{42 scaling}. The continuous lines are solutions derived via the T-matrix formalism, while the gray dashed line is the Mie solution computed for a spherical particle with the same volume and composition as the model. At short wavelengths, the use of a fixed truncation order (blue line) leads to convergence problems, which are solved by the introduction of dynamic orders (red) and separate inner and outer order assignments (black). On the other hand, all models converge to a Rayleigh regime at long wavelengths.
\label{fig_dyn_orders}}
\end{center}
\end{figure}
An example of the \texttt{np\_cluster} performance benchmarking tests is shown in Fig.~\ref{fig_np_cluster_bench}, while a description of the relevant characteristics of the benchmarking models is given in Table~\ref{tab_models}. In terms of performance, the adopted parallel implementation is able to grant an execution time reduction by a factor larger than $10$, when running on commercial hardware. The same code implementation shows an overall speed-up factor larger than $20$, when executed on a single node of a computing farm, with a nearly linear improvement of the scaling, if more nodes are used simultaneously. The linearity of the scale is easily understood as a consequence of the embarrassingly parallel nature of the model calculation on different wavelengths, with a minor overhead (typically in the order of few seconds) to distribute the calculation data via MPI messages and to arrange for the underlying thread hierarchy that addresses non-embarrassingly parallel code sections via work-sharing tasks.

The accuracy of the results is tested in every case for which we can recover the results of the original implementation by comparing the full code output, while we provide Mie solutions of equivalent spheres, for the cases that cannot be solved under the limitations of the original implementation, knowing that both the T-matrix predictions and the Mie solutions converge to the Rayleigh regime when the radiation wavelength becomes much larger than the particle size. Such a consistency check is illustrated, for the case of model \textit{42 scaling}, in Fig.~\ref{fig_dyn_orders}. From an inspection of the results, we see how all the calculations converge to the same limit at long wavelengths, while showing substantial differences at short ones. Such dramatic differences arise from the strong dependency of the radiation interaction effects on particle structural details that have nearly the same size as the radiation wavelength. In this case, the choice of a proper multipole expansion truncation order acquires a critical role. Indeed, the use of a fixed order that is too low to solve the particle as a whole (like $l_{max} = 10$, in this case) can lead to lack of convergence at the shortest wavelength scales. Using a higher order, such as $l_{max} = 14$, on the other hand, increases the complexity of the calculations and it may lead to numerical stability problems in the long wavelength regime. Instead, our results suggest that the use of maximum order separation, adopting $l_{max} = l_I = 6$, when dealing with the single monomers, and $l_{max} = l_E = 15$, for the particle as a whole, in combination with dynamic truncation order assignment, grants for a convergent solution with the best possible performance and prevents the occurrence of numerical instability.

\subsection{\texttt{np\_trapping} benchmarks}

To estimate the code performance for particle trapping calculations, we used a similar approach to the one introduced in the previous case, which involves running the calculation of the same models with the original code implementation and then comparing the results with those produced by the parallel implementation of \texttt{np\_trapping}, executed on different hardware configurations. Here the benchmarking model evaluates the trapping conditions of a spherical plastic particle with radius $r_{sph} = 200\,$nm, refractive index $n_{sph} = 1.7$, immersed in water ($n = 1.33$) and exposed to a TEM$_{00}$ Gaussian laser beam, with wavelength $\lambda =785\,$nm, aligned with the $z$-axis direction and linearly polarized along the $x$-axis direction \citep[a detailed description of the instrumental setup is given by][]{Rezaei24}. The full output of the calculation consists of the forces and torques exerted on the particle in a grid of $51 \times 51 \times 51$ equally spaced points, distributed around the beam focus with a grid step of $71\,$nm.

\begin{figure}[t]
\begin{center}
\includegraphics[width=0.455\textwidth]{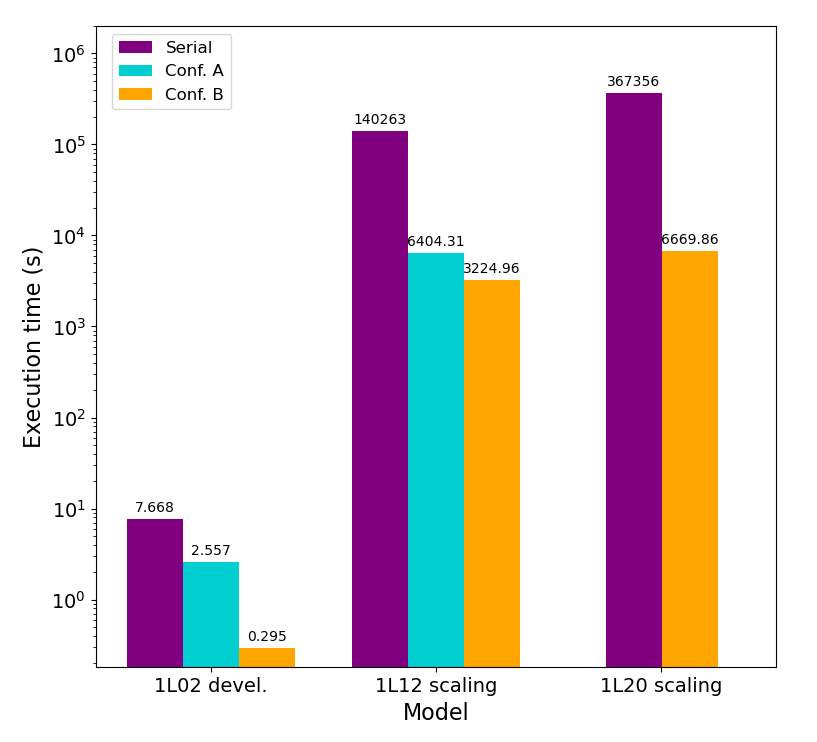}
\caption{Comparison of \texttt{np\_trapping} execution times for a development test case and two scaling cases, solving, respectively, for the trapping of a spherical particle in an axial radiation beam with multipole truncation order set at $l_{max} = 2$, $l_{max} = 12$ and $l_{max} = 20$. The solution of this last case was not profiled on Configuration A. \label{fig_np_trapping_bench}}
\end{center}
\end{figure}
The solution of the trapping problem can be regarded as a two-step process. The first and most demanding stage is the calculation of the radiation field originated by an input beam (usually a collimated laboratory laser beam), computed on a grid of points encompassing the sample particle trapping volume. The second step, instead, uses the particle T-matrix representation at the relevant wavelength to calculate the forces and torques that the particle would experience if placed in each vertex of the pre-computed radiation field grid. The advantage of this subdivision is that the trapping solution of different particle models for the same radiation beam can be effectively extracted by first computing the radiation field grid and then using the same solution in combination with all relevant particle models.

\begin{figure*}[t]
\begin{center}
\includegraphics[width=0.95\textwidth]{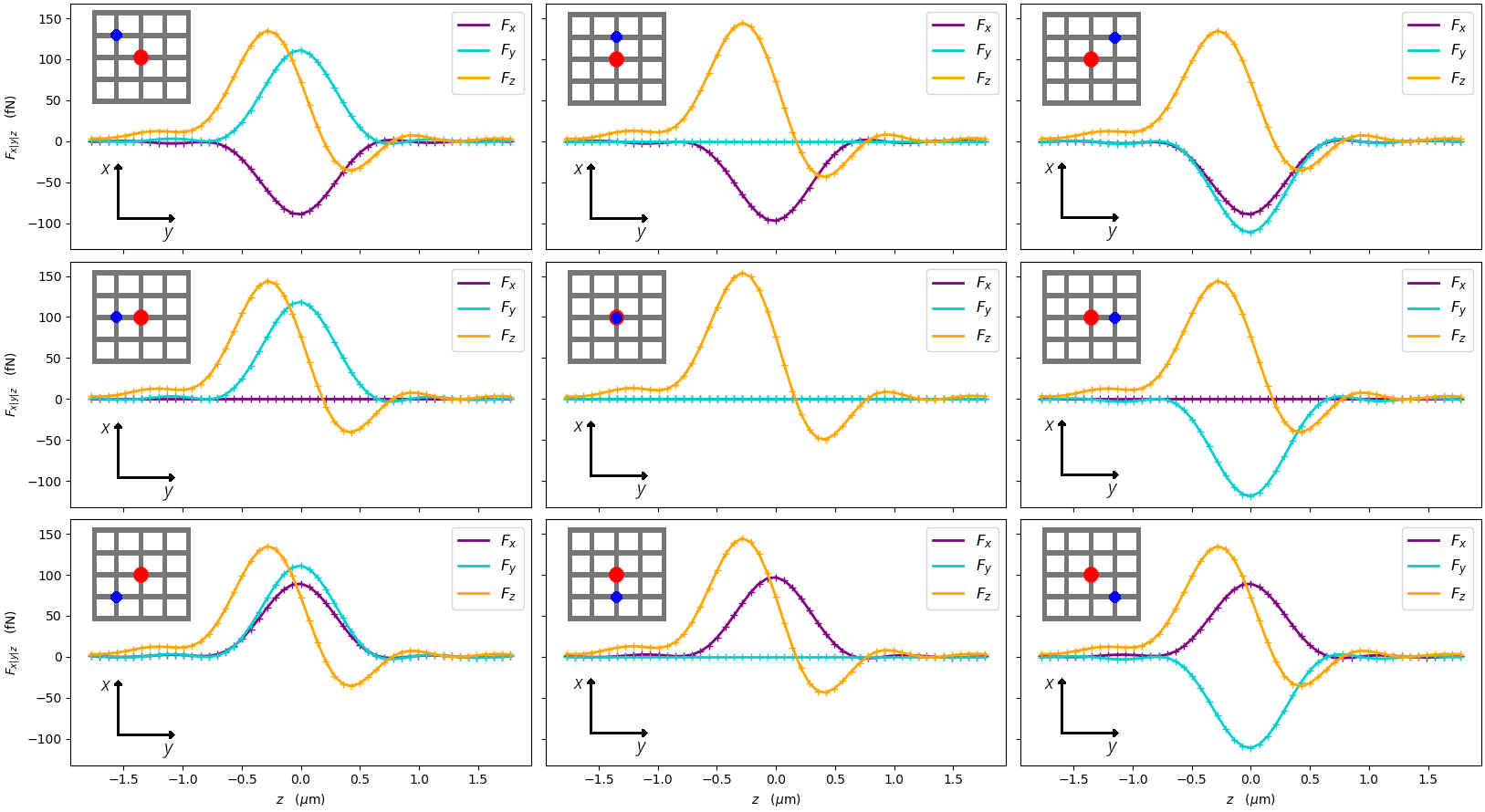}
\caption{Components of the radiation forces exerted on a spherical particle by an axial laser beam aligned with the $z$ direction as a function of position on the $z$-axis, for a particle located on the beam focal axis (central plot), or on the nearest $(x,y)$ grid points (surrounding plots). The particle position is represented as a blue circle with respect to the focal point (red circle) in each panel. The grid spacing is $71\,$nm. Continuous lines represent the results of \texttt{np\_trapping}, while the \textit{plus} signs are the results of the original serial implementation. For this geometry, only components along the $z$-axis are expected to be non-zero on the focal axis. Positive values of $F_x$ push the particle upwards. Positive values of $F_y$ push the particle to the rightwards. Positive values of $F_z$ push the particle away from reader. \label{fig_rad_forces}}
\end{center}
\end{figure*}
In the case of \texttt{np\_trapping} the parallelization strategy focused on computing the radiation field simultaneously on multiple grid vertices. Fig.~\ref{fig_np_trapping_bench} shows the comparison of the execution times needed to solve the case of a spherical particle in an axial laser beam, whose multipole order expansion was truncated at $l_{max} = 2$, for development purposes, then at $l_{max} = 12$ and $l_{max} = 20$ for scaling purposes. The accuracy of the results was tested by comparing all the numerical output of the various code implementations, to assess that no differences beyond the effects of numeric noise on negligible values occur, and by extracting diagnostic plots, such as, for instance, the magnitude of the radiation force components around the beam axis, which, in this particular case, are expected to trap the particle close to the focal point, as shown in Fig.~\ref{fig_rad_forces}. The results of our tests show that accurate solutions of the trapping problem can be obtained in the framework of dynamic order selection, which allows for substantial performance improvement on different hardware architectures, while still preserving the necessary accuracy of results.

\section{Discussion}
\label{sec-discussion}

The ability to produce accurate extinction models for non-spherical particles, as well as the possibility to compute the dynamical effects of the interaction between particle and radiation beams (i.e. optical tweezers), has fundamental implications in the study of suspended particles embedded in transmissive media and in laboratory applications involving the control of samples via laser beams. Using various types of computing approaches, many studies agree on the result that deviation from the simple spherical symmetry leads to substantial effects on the efficiency of particles with the same composition to intercept radiation \citep{Saija03b, Lodge24, LaMura25}. Such dramatic effects are evident even from a simple inspection of the extinction cross-sections predicted in our development and scaling test models. For this reason, a proper exploration of the role played by the particle structural properties and size distribution is a critical ingredient to perform quantitative studies based on the analysis of transmission spectra.

In principle, the T-matrix formalism is the most suitable approach to deal with the simulation of particles with arbitrary shape and composition, thanks to its ability to use different types of particle constituents. This aspect is particularly relevant for realistic models that aim at reproducing the properties of particles that are exposed to a variety of physical conditions during their formation process. Some relevant examples in astrophysical applications involve the study of solid particles forming in different layers of planetary atmospheres, but also the formation and evolution of dust grains in the interstellar medium and in proto-planetary disks. In all these cases, the ability to model the particle as an aggregate of potentially very different components, is a fundamental ingredient, which is inherently supported by T-matrix based approaches. Calculations based on other methods, such as DDA, can also address the same type of problems, but they need to be fully solved for every possible combination of incident and scattered radiation directions, limiting the possibility to accurately model a population of randomly oriented particles.

\begin{table*}[t]
\begin{center}
\caption{Summary of the model parameters used to benchmark \nptm\ applications and to test their performance and accuracy with respect to the original serial implementation. The table columns report, respectively, the name used to refer to the model, the number of spherical units composing the particle, the minimum and the maximum radius of the spherical units (these values are identical if only one sphere type is used), the inner and the outer multipole field truncation order, the size of the associated T-matrix, the minimum and the maximum wavelengths covered by the calculation, the number of computed wavelengths, the materials of the spherical monomers and the number of differential directions or spatial grid points extracted$^{(a)}$. \label{tab_models}}
\begin{tabular}{cccccccccccc}
    \hline
    \hline
    Model name & $n_{sph}$ & $r_{sph}^{min}$ & $r_{sph}^{max}$ & $l_i$ & $l_e$ & T-matrix size & $\lambda_{min}$ & $\lambda_{max}$ & $N_\lambda$ & Matrial(s) & Diff. \\
    \hline
    48 devel. & $48$ & $5.0\,$nm & $5.0\,$nm & $2$ & $2$ & $[768 \times 768]$ & $420\,$nm & $600\,$nm & 181 & porphyrin & $1$ \\
    4 scaling & $4$ & $20.0\,$nm & $200.0\,$nm & $12$ & $12$ & $[1344 \times 1344]$ & $400\,$nm & $800\,$nm & 401 & plastic, gold & $25$ \\
    16 scaling & $16$ & $100.0\,$nm & $100.0\,$nm & dyn. & $16$ & $[9216 \times 9216]$ & $0.2\, \mu$m & $25.0\, \mu$m & 76 & enstatite & $1$ \\
    42 scaling & $42$ & $34.3\,$nm & $68.6\,$nm & $6$ & $15$ & $[4032 \times 4032]$ & $100\,$nm & $1\, \mu$m & 181 & silicates, carbon & $1$ \\
    1L02 devel. & $1$ & $200.0\,$nm & $200.0\,$nm & $2$ & $2$ & $[16 \times 16]$ & $785\,$nm & $785\,$nm & 1 & plastic & $51^3$ \\
    1L12 scaling & $1$ & $200.0\,$nm & $200.0\,$nm & $12$ & $12$ & $[336 \times 336]$ & $785\,$nm & $785\,$nm & 1 & plastic & $51^3$ \\
    1L20 scaling & $1$ & $200.0\,$nm & $200.0\,$nm & $20$ & $20$ & $[880 \times 880]$ & $785\,$nm & $785\,$nm & 1 & plastic & $51^3$ \\
    \hline
\end{tabular}
\end{center}
\begin{footnotesize}
$^{(a)}$ The number of differential calculations equals the number of scattering directions computed by \texttt{np\_cluster} or the number of spatial grid points solved by \texttt{np\_trapping}.
\end{footnotesize}
\end{table*}
In this framework, \nptm\ represents a parallel and scalable solution, designed to improve the performance of models based on the T-matrix formalism, supporting execution both on single work-stations and large scale computing facilities. The test cases presented in this work illustrate the performance gains obtained while running a sequence of models on increasingly powerful hardware systems. In addition to the parallel code structure, which takes full advantage from the multi-core processing capabilities of modern hardware systems, \nptm\ offers a set of auxiliary functions that improve the algorithmic stability and simplify the code use and its integration with data analysis and visualization software. Indeed, the introduction of features like the dynamic truncation order calculation and the iterative refinement greatly expand the range of wavelengths that can be reliably covered. Moreover, while all the calculation results can still be optionally produced in a legacy format (a feature that is mainly used for consistency tests with other implementations), more efficient data structures have been introduced to support output in optimized standard formats, such as \texttt{HDF5}, as well as to define input model parameters via human readable configuration files written according to the \texttt{YAML} standard. Such functionality is offered in the form of \texttt{python} scripts designed to help in the model definition and in the inspection of results with minimal dependencies on external libraries.

\begin{figure*}[t]
    \begin{center}
        \includegraphics[width=0.95\textwidth]{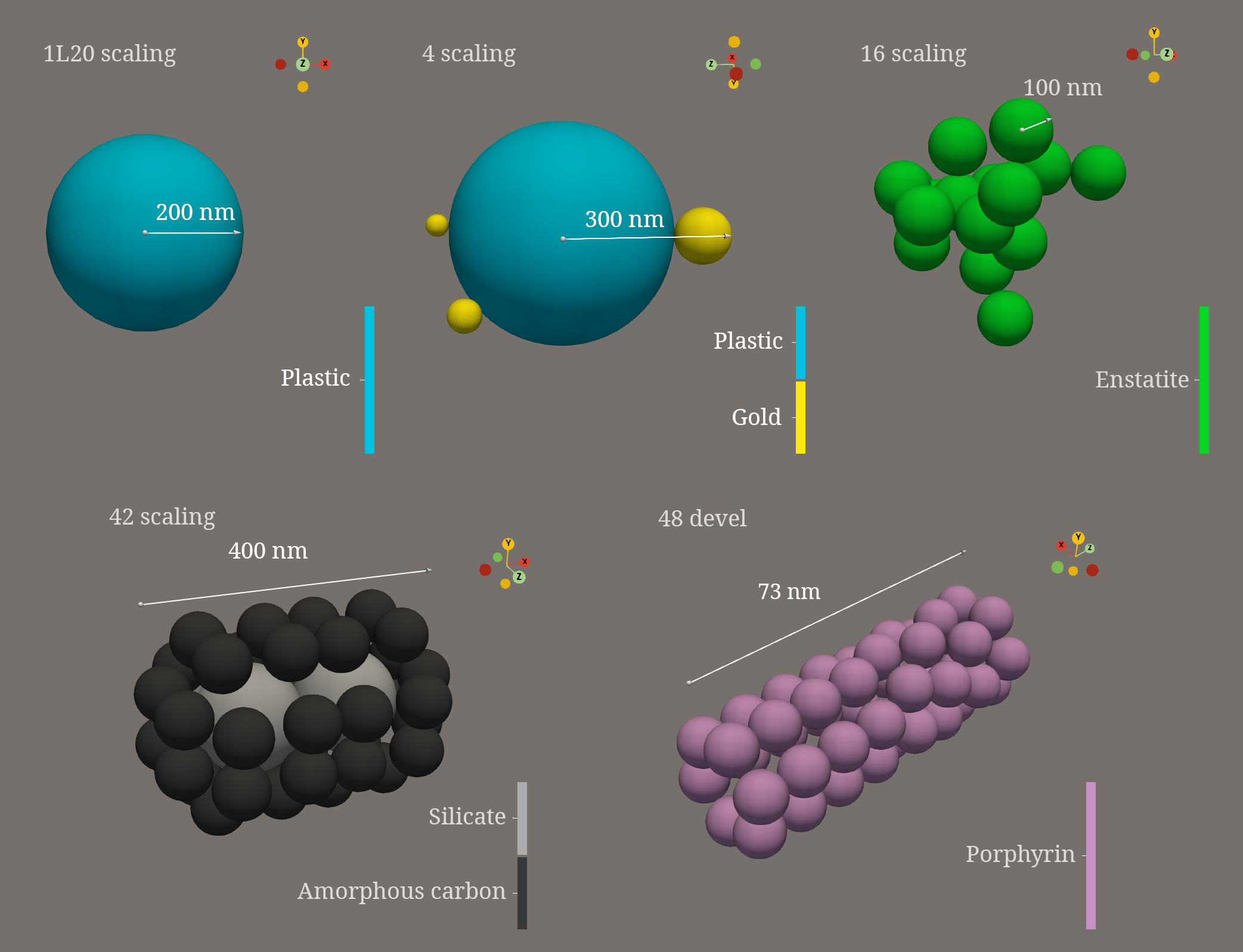}
        \caption{3D representation of the models used to benchmark the code. \label{fig_appendix_b}}
    \end{center}
\end{figure*}

A schematic representation of the models used to benchmark the code is summarized in Fig.~\ref{fig_appendix_b}, as appendix.

\section{Conclusions}
\label{sec-conclusions}

In this work we presented \nptm, the new parallel implementation of the T-matrix formalism. Although the code, in its present form, is mainly focusing on the calculation of the extinction cross-sections and the dynamic effects of the interaction of radiation with particles characterized by arbitrary composition and chemical properties, the T-matrix solution provides the grounds to describe a much broader range of physical processes. Being formally able to describe the radiation field within the particle, processes such as fluorescence, self-absorption, thermal balance and anisotropic emission stresses can be accurately modeled.

\nptm\ offers an enhanced possibility to run model calculations based on the T-matrix formalism. The advantages over the original implementation used as a reference for the parallel solution described here are not exclusively limited to improved performance, but they also include a better management of the available hardware resources and a more intuitive model configuration process. Documented scripts to configure the models and to parse the results relieve the user from the burden of editing potentially large machine readable datasets and help in the result inspection stage, allowing for conversion of the calculation outcomes to standard data representation formats. While a full set of model calculations, able to describe the properties of realistic distributions of particles in astrophysical environments, is still ongoing, we expect that the \nptm\ implementation of the T-matrix formalism will bring a relevant contribution to the possibility of modeling radiation scattering and extinction processes with an unprecedented degree of accuracy in several fields of astrophysical investigation and laboratory experiments.

\section*{Acknowledgments}

Supported by Italian Research Center on High Performance Computing Big Data and Quantum Computing (ICSC), project funded by European Union - NextGenerationEU - and National Recovery and Resilience Plan (NRRP) - Mission 4 Component 2 within the activities of Spoke 3 (Astrophysics and Cosmos Observations) and by the Italian Ministry of University and Research (MUR), through PRIN Exo-CASH (PRIN 2022 project no. 2022J7ZFRA). This work was completed in part at the CINECA GPU HACKATHON 2024, part of the Open Hackathons program. The authors would like to acknowledge OpenACC-Standard.org for their support. The authors gratefully thank the journal reviewers for their useful discussion and suggestions. The authors also acknowledge the relevance of the ESA Ariel Mission to this study, whose results are intended to be integrated into analysis frameworks employed by mission collaborators. The \texttt{NP\_TMcode} software is the parallel implementation of the T-matrix formalism, currently under development at INAF - Astronomical Observatory of Cagliari, on the basis of the codes originally written by F. Borghese, P. Denti and R. Saija. It is registered in the Astrophysics Source Code Library (ASCL) as \texttt{ascl:2510.003} (\url{https://ascl.net/2510.003}).

\bibliographystyle{elsarticle-harv} 
\bibliography{ms}

\begin{figure}[t]
    \begin{center}
        \includegraphics[width=0.45\textwidth]{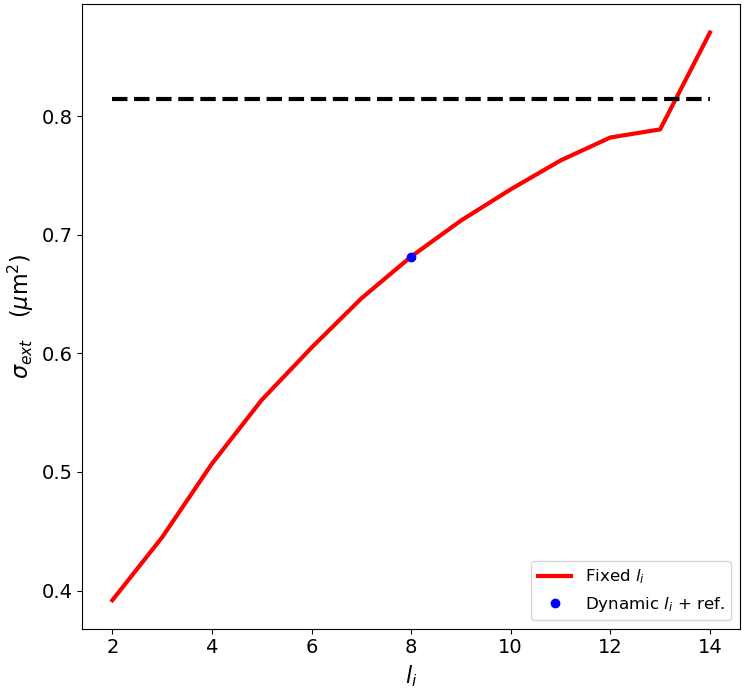}
        \caption{Extinction cross-section computed with various internal orders at $\lambda = 2.11 \mu$m for the same model shown in Fig.~\ref{fig_instability}, but using iron as particle material. As in the right panel of Fig.~\ref{fig_instability}, the blue point is the extinction cross-section $\sigma_{ext}$ at $\lambda = 2.11\, \mu\mathrm{m}$ computed in dynamic order, the red curve is computed for different values of fixed $l_i$, and the black dashed line represents the convergence limit. \label{fig_appendix_a}}
    \end{center}
\end{figure}
\section*{APPENDIX A}
\subsection*{Numerical stability for conductive materials}
The problem of numerical stability is connected with the dynamic range of the elements of the T-matrix. In case of materials with a complex refractive index dominated by the imaginary part, such as iron or other effective conductors interacting with long wavelength radiation, the structure of the T-matrix is such that optimized inversion algorithms become unstable before actual convergence is achieved, as shown in Fig.~\ref{fig_appendix_a}.

In circumstances when the particle is dominated by conductive material, reasonable solutions can still be obtained using less optimized, but more numerically stable, matrix inversion algorithms, such as the internal $LU$ factorization, at the cost of a calculation speed decrease of a factor of $8$.

\end{document}